# The value creation potential of digital humans

Corresponding author: Araz Zirar, PhD

Department of Management
Huddersfield Business School
University of Huddersfield
Huddersfield, UK
Tel: +44 (0)1484 47(2608)
Email: A.Zirar@hud.ac.uk
ORCID: 0000-0001-6974-2991
LinkedIn: https://www.linkedin.com/in/arazzirar/

**Funding details.** Not applicable

**Disclosure statement.** The author declares that he has no conflict of interest.

**Data availability statement.** The search string, the list of articles, the search protocol file can be provided upon request.

# The value creation potential of digital humans

**Abstract**

'Digital humans' are digital reproductions of humans powered by artificial intelligence (AI) and capable of communicating and forming emotional bonds. The value creation potential of digital humans is overlooked due to the limitations of digital human technologies. This article explores the value creation potential and the value realisation limitations of digital humans. The analysis is based on a review of 62 articles retrieved from the Web of Science database. The analysis suggests that digital humans have the potential to alleviate labour/skill shortages, reduce the natural human element in high-risk tasks, avoid design errors, improve the ergonomics of products and workplaces, and provide guidance and emotional support, all of which will benefit natural humans in the workplace. However, technical limits, evolving understanding of digital humans, the social significance and acceptance of digital humans, ethical considerations, and the adjustment of legal tradition limit the value realisation. This review suggests that digital humans' perceived usefulness and ease of development determine organisations' willingness to utilise this technology. Overcoming the limitations, which still involve engineering challenges and a change in how they are perceived, will positively affect realising the value potential of digital humans in organisations.

# 1. Introduction

'Digital humans' are visually life-like digital reproductions of humans capable of interacting, responding to requests, and completing tasks (Silva & Bonetti, 2021). In contrast to interactions between entities with no personality, such as the digital twin of a physical inanimate object, a digital human is expected to display individuality and engage in interactions based on the diversity that emerges from such individuality (Guadagno et al., 2011). A digital human is also expected to reproduce social characteristics such as human behaviour and communication in digital space (Guadagno et al., 2011). As a result, they are expected to blink, wink, frown, nod, and smile and have personality and character, allowing them to converse face-to-face with humans (Moore et al., 2022; Silva & Bonetti, 2021). Since digital humans are digital data that can be copied, merged, and transmitted, they are projected to improve human competence and decision-making (dos Santos et al., 2021) and build group intelligence (Yablochnikov et al., 2021).

While there are still technical limitations to what digital humans can do (Baptista et al., 2020), it is projected that digital humans will be able to perform human-only tasks (Holford, 2019). Therefore, digital humans are projected to be part of an organisation's 'intelligent augmentation layer' to supplement or replace human labour (Baptista et al., 2020).

Will humans still be needed to perform manual or repetitive labour in the workplace, such as scheduling tasks, even if they prefer it? It is reasonable to foresee that the employment of digital humans in the workplace will impact human work, reshaping the digital/physical work configuration/distribution (Baptista et al., 2020). Although research is still needed in this area, one observation is that using digital humans requires reorganising digital/human labour, which is critical in defining modern organisations (Baptista et al., 2020). Moreover, the employment of digital humans in a workplace goes beyond the visible labour requirements to how they fundamentally transform an organisation (Baptista et al., 2020).

On the other hand, the intricacy of recognising a workforce's human element adds to the complexity of developing digital humans. The value creation potential of digital humans is often overlooked due to the limitations of existing digital human technologies. This review explores the value propositions of digital humans, offering insights into value creation potential and limitations that impede value realisation.

A systematic search was conducted to locate and retrieve literature on 'digital humans' (Tranfield et al., 2003). After applying the inclusion and exclusion criteria, the 62 returned articles were thematically analysed following Braun & Clarke's (2006) thematic approach.

This review has five sections. The concept of 'digital humans' is defined in the second section. The third section describes the literature search, which includes the research approach, the search string, the exclusion and inclusion criteria, and the database selection. In section four, the thematic analysis is presented. The review's future directions, conclusion, contribution, and limitations are presented in the fifth section.

## 2. Digital humans

'Digital humans,' or the life-like reproduction of people in cyberspace, is still a technological breakthrough in the future (Holford, 2019; Silva & Bonetti, 2021). While progress has been made in creating human-like digital beings in cyberspace, such advancements are still in their infancy and continue to be engineering challenges (Demirel et al., 2021; Yam et al., 2020). Existing digital human prototypes have been trained to do human functions, including assisting and providing emotional support (see https://digitalhumans.com/ for more information).

In this review, a 'digital human' is a life-like entity powered by artificial intelligence (AI) capable of conversing, communicating, and forming an emotional connection like a human (Guadagno et al., 2011; Silva & Bonetti, 2021). This definition elaborates on the several characteristics a digital being must possess to be a digital human. First, a digital being must be a 'life-like being' and resemble a human. Second, artificial intelligence technology must power this digital entity to converse, interact and form emotional bonds.

The existing examples of digital humans can participate in human-like activities like conversing and communicating utilising artificial intelligence technologies. As per a study (Guadagno et al., 2011), they can also produce a digitally rendered smile that humans positively receive. However, this technology is still in the future, and digital humans' functionality is still evolving (Demirel et al., 2021; Silva & Bonetti, 2021).

## 3. Research method

The review process was organised around the question: What is the value proposition of digital humans in organisations? The question guided the selection of keywords and the search, organisation, analysis, and synthesis of relevant literature (Rousseau et al., 2008).

A manual search was performed in Google Scholar using the terms ('digital twin' AND human) to accumulate relevant keywords. Then, using a combination of the accumulated keywords, a search was performed in the Web of Science electronic database. The search string (before applying the database inclusion and exclusion criteria) was: **TI** OR **AB** OR **AK=( "digital human" OR "virtual human" OR ( ( "mixed reality" OR "digital twin" OR "cyber-physical system" OR "computational intelligence" OR "autonomous systems" OR "digital assistant" ) AND human ) )**. The field codes TI, AB, and AK were used in WoS to request the database to search the Title, Abstract, and Keywords of documents.

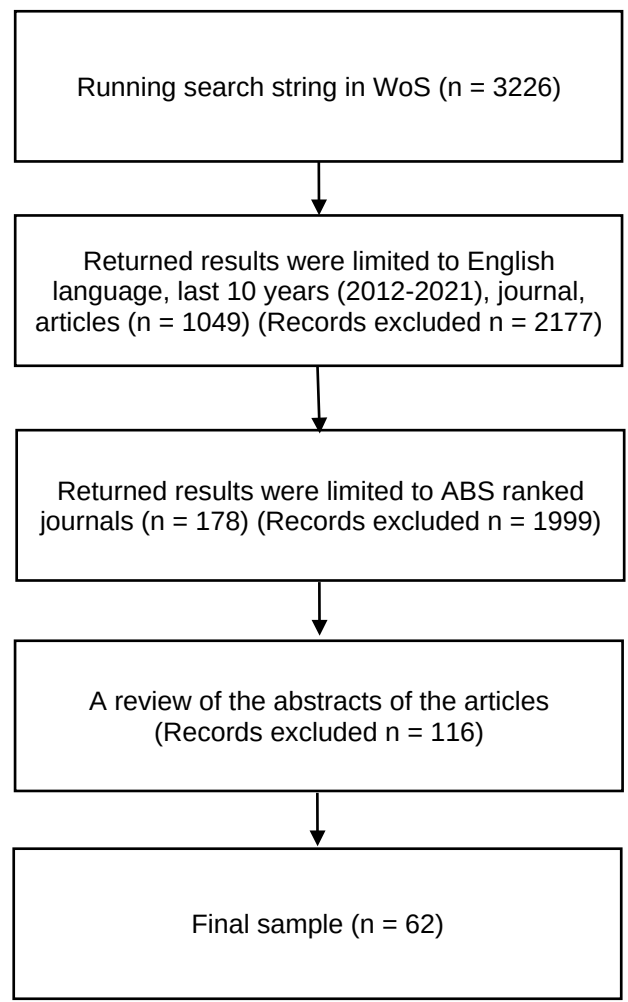

Figure 1: The process of selecting the articles for this review

Because of the nature of the topic, which entails a steady advancement in technology, the depth of publications was prioritised over the breadth. The WoS database provides this, as it archives more extended periods of peer-reviewed journals published by major publishing houses (Zhu & Liu, 2020). Following the exclusion and inclusion criteria (Figure 1), 62 articles were retrieved from the WoS database.

The objective of the data analysis stage was to break down the accumulated data into smaller pieces to understand the selected list of articles (Tranfield et al., 2003). The data were subsequently organised into themes (Braun & Clarke, 2006).

This review was not intended to generalise the findings but to provide insights into digital humans' value propositions. Thematic analysis, therefore, was interpretive, intending to identify emerging themes and highlight links (Braun & Clarke, 2006). Themes were linked iteratively, meaning that the analysis process entailed forward and backward movement in terms of searching, refining, and reporting (Braun & Clarke, 2006).

## 4. Findings and discussion

Digital humans add value to businesses. However, the value propositions have yet to be realised due to the limitations of digital human technologies. This section delves into value proposition themes before highlighting the limitations of digital humans.

*4.2 Value propositions of digital humans*

**To reduce worker fatigue and discomfort.** In optimising tasks, digital humans have been employed in proactive ergonomics to reduce fatigue and discomfort in worker tasks (Chander & Cavatorta, 2018; Maurice et al., 2017). Minimising poor ergonomic postures and corresponding body part discomforts in waste sorting tasks (Emmatty et al., 2021; Lämkull et al., 2009; Sanjog et al., 2019), improving visibility for delivery drivers (Choi et al., 2009; Potvin & Potvin, 2019; Summerskill et al.,

2016), reducing the length between paramedics and their equipment bags (Harari et al., 2020), and virtually studying work-related risk factors in plastic furniture manufacturing factories (Mazzola et al., 2017; Sanjog et al., 2019) are examples. These efforts aim to reduce worker fatigue and discomfort by virtually analysing fatigue resistance and discomfort of muscle groups using digital humans (L. Ma et al., 2011; Y. Ma et al., 2011; Zhang et al., 2010).

Unlike traditional ergonomic practices that rely on limited user data, ergonomic recommendations, or designer experience (Harih & Dolšak, 2013; Vyavahare & Kallurkar, 2016), using digital humans allows researchers to evaluate worker fatigue and discomfort with specific tools in a virtual setting while also obtaining quantitative results (Harih et al., 2021; Harih & Dolšak, 2013; Vyavahare & Kallurkar, 2016). While humans can understand abstract concepts in this context, the precision, efficiency, and capability of using digital humans in proactive ergonomic decisions is unmatched (Chu et al., 2006; Ghosh, 2002). These efforts have been effective (Lämkull et al., 2009), and they improved ergonomics for manufacturing line workers (Lämkull et al., 2009).

However, such ergonomic improvements were made in standing and unrestricted working positions (Lämkull et al., 2009). This could indicate that the use of digital humans is still in its early stages and that digital humans can only be used to enhance simple design aspects. While this suggests that humans are no longer required to consider proactive ergonomics design decisions, such as the optimal placement of auxiliary equipment at a workplace, humans should still make other sophisticated design decisions.

Further, using digital humans in this context might suggest a collaboration between humans and digital humans to reduce fatigue and discomfort for humans.

Although humans continue to use digital humans in a simulated environment, it is unclear whether digital humans will in the future act as real counterparts to humans in this collaborative endeavour (Holford, 2019).

**To evaluate product and workplace.** Manual labour is still common, and humans undertake manual labour (Giudice et al., 2021). Even outside the office, there are still manual tasks that human workers choose to do, such as picking fruit. Hiring humans to perform them is likely less expensive (Harih & Dolšak, 2013). Ensuring the right equipment or workplace design ensures that manual labour does not create health issues for workers, such as musculoskeletal illnesses (Harih & Dolšak, 2013).

Digital humans are utilised to evaluate products and workplace designs (Chander & Cavatorta, 2018; Jung et al., 2009). Using digital humans facilitates the optimal design of tools and workspace layouts (Harih & Dolšak, 2013). However, unlike employing humans, the use of digital humans in the design process does not have to be iterative (Hanson et al., 2006; Harih & Dolšak, 2013; Lämkull et al., 2009; Summerskill et al., 2016).

Unlike the traditional approach to product and workplace design, which involves real-world evaluation, such evaluation can be carried out in a virtual environment without the physical product or workplace (Chander & Cavatorta, 2018; Emmatty et al., 2021; Jung et al., 2009). Unlimited design possibilities can be considered in a virtual world (Choi et al., 2009; Chu et al., 2006).

Unlike humans, the speed, efficiency, and capability of digital humans, as well as their flexibility to be copied as desired, enables the simulation of design considerations without the need to create such designs in the real world (Chu et al., 2006; Ghosh, 2002; Sanjog et al., 2019). For example, creating a wheelchair to improve the mobility of a disabled worker in a work environment is possible by utilising digital humans in a virtual environment (Chateauroux & Wang, 2010; Lim et al., 2004; Xiang et al., 2018).

While optimal designs benefit workers, they benefit businesses as well. For instance, businesses can retain their workers, particularly when a labour shortage exists in a particular industry, such as the care industry. Therefore, using digital humans aids in evaluating tools and workspace designs (Harih & Dolšak, 2013). While using digital humans to evaluate products and workplaces can be convenient, efficient, and effective (Kuo & Wang, 2009), measurement errors in constructing digital humans reduce their value in this context (Gragg et al., 2013; Ozsoy et al., 2015).

**To avoid design errors.** Digital humans are used to avoid design errors (Caputo et al., 2019). These include identifying design factors that influence visibility (Choi et al., 2009), creating models (Harih et al., 2021), optimising the position of equipment bags (Harari et al., 2020), specific human-centred design in manufacturing (Sanjog et al., 2019), creating products (Xiang et al., 2018), determining fatigue resistances of muscle groups (L. Ma et al., 2011), evaluating work layout for tasks (Emmatty et al., 2021; Mazzola et al., 2017), considering the optimal shape of tools (Harih & Dolšak, 2013), analysing and visualising interiors (Hanson et al., 2006), understanding blind spots in the vision of drivers (Summerskill et al., 2016), and investigating the virtual teacher's positions to improve learning outcomes in mixed-reality situations (Nawahdah & Inoue, 2013).

Further, humans will be involved in a physical production line in various ways, yet design decisions will determine their workplace experience. The rationale is that correcting a design error in industries such as manufacturing can be costly, especially at the production stage (Caputo et al., 2019; Emmatty et al., 2021; Gragg et al., 2013; Hu et al., 2011).

In this context, digital twins will include the development of digital humans during the design phase to improve the design decisions of a production line (Caputo et al., 2019; Emmatty et al., 2021; Gragg et al., 2013; Hu et al., 2011). Any design errors must be addressed in the design phase, and using a digital twin allows an organisation to do so (Caputo et al., 2019; Emmatty et al., 2021; Gragg et al., 2013; Hu et al., 2011). Therefore, digital twins will extend to identifying design issues using digital humans (Ozsoy et al., 2015).

**To alleviate labour/skill shortage.** There is a shortage of nurses (Rafferty, 2018), and digital humans can be a potential solution (Tan et al., 2021). One advantage of digital humans over humans in this context is that they can be duplicated as needed (Tan et al., 2021; Xiang et al., 2018). Digital humans' narrow intelligence makes them cost-effective for addressing specific skill shortages, such as having a meaningful conversation with a care home resident.

However, while using digital humans may alleviate labour/skill shortages, it may have negative job implications (Tan et al., 2021). Something that has the potential to alleviate labour/skill shortages could inadvertently increase unemployment (Tan et al., 2021). This is primarily due to the nature of digital humans. They are digital and can be reproduced as needed (dos Santos et al., 2021). Negative job implications can cause conflict with digital humans and societal consequences, with unemployed people expressing dissatisfaction by causing civil unrest. Moreover, the development of such digital humans is still in its inception

(Gelbrich et al., 2021). If they care for vulnerable people and provide psychological therapy, for example, the evolving nature of the technology may result in an everlasting distrust of their abilities (Tan et al., 2021).

**To reduce the human element in operations.** Digital humans are projected to reduce the human element in operations (Jiang et al., 2015; Silva & Bonetti, 2021). These projections are mainly driven by the assumption that defects and waste in operations are generated by human behaviour (Jagdev et al., 2008; Jiang et al., 2015; Kim & Kim, 2020). For example, searching for specific information by humans is resource-intensive when compared to digital humans (Hanson et al., 2006; Jiang et al., 2015), or design processes that involve humans are iterative while this is not required when using digital humans (Gragg et al., 2013; Harih & Dolšak, 2013; Munoz et al., 2016). Further, industries such as fashion cannot survive a pandemic such as COVID-19 when humans are involved in operations (Fan et al., 2021; Leng et al., 2021; Silva & Bonetti, 2021).

     However, this line of research acknowledges that customers prefer more than gesture, text and speech interactions with digital humans; they prefer such interactions to extend to include, for instance, comprehension of vague language (Kuo & Wang, 2009; Large et al., 2017; Silva & Bonetti, 2021). Therefore, while there have been projections to use digital humans to reduce the human element in operations, we are yet to fully assess, understand, and apply individual differences in adaptation to the use of digital humans (Kim & Kim, 2020; Matthews et al., 2021).

     Moreover, while digital humans are projected to reduce the human element in operations (Jagdev et al., 2008; Kim & Kim, 2020), costly skilled human resources are required to develop them (Lee et al., 2011). Further, the human element in

decision-making is still crucial, especially when a decision involves intuition (Jaiswal et al., 2021).

In this context, however, the main argument in favour of digital humans is that humans' reliability is unpredictable, and as a result, humans under pressure may behave unpredictably (Jiang et al., 2015). On the other hand, digital humans are constructed using predetermined data, making them predictable (Lämkull et al., 2009).

Therefore, organisations can rely less on the human workforce and use digital humans. This argument extends beyond reducing a business's labour costs to protecting human health and safety during a pandemic (Silva & Bonetti, 2021). On the other hand, customers may not tolerate digital humans' technical limitations, such as only providing limited voice or chat interaction (Silva & Bonetti, 2021).

Research (Khalid et al., 2018; Wilson & Daugherty, 2019) suggests that digital and natural humans should collaborate instead of replacing each other. In this symbiotic relationship, natural humans interact with digital humans to augment their abilities (Khalid et al., 2018). Digital humans can handle repetitive inquiries, while natural humans handle nuanced inquiries (Wilson & Daugherty, 2019). However, a clear distinction between tasks to be performed by natural humans and those to be performed by digital humans is still evolving.

**To reduce cost and save time on product development.** While developing digital humans is costly and complex (Lee et al., 2011; Ranger et al., 2019), businesses can use them to save time and cost when developing products used by humans (Hanson et al., 2006; Ozsoy et al., 2015). Businesses that use digital humans to save time and cost on product development must make the initial costly and time-consuming investment in their development. However, the return on investment is uncertain because digital human technology is still in its formative stages.

Furthermore, prototyping with digital humans appears to have reduced costs associated with correcting design errors and time-to-market products (Caputo et al., 2019; Xiang et al., 2018). Prototyping an optimal product design (Xiang et al., 2018), employing digital assistants to understand navigational needs (Large et al., 2017), and analysing and visualising interiors in connection to human characteristics are a few examples (Hanson et al., 2006; Munoz et al., 2016).

The cost/time benefits primarily come from reducing the requirement for in-person human data collection and physical prototyping in hazardous work conditions (Demirel et al., 2016). Identifying work-related risk factors and devising specific human interventions, particularly in high-risk work environments such as plastic furniture manufacturing plants, have saved businesses time and cost (L. Ma et al., 2011; Sanjog et al., 2019).

However, the initial time/cost investment is significant since developing digital humans requires the use of skilled human resources (Lee et al., 2011; Ranger et al., 2019), and such investment is required until the humans are no longer involved (Jagdev et al., 2008).

**To provide emotional support.** One might argue that humans are inconsistent in their emotional support. This is where the argument for using digital humans to provide emotional support comes into play (Gelbrich et al., 2021; Guadagno et al., 2011). If programmed, digital humans' emotional support, such as reassuring expressions, will always be consistent and accompanied by a 'perfect' smile (Gelbrich et al., 2021; Guadagno et al., 2011).

Individuals who have failed a task can receive emotional support from digital humans through reassuring expressions (Gelbrich et al., 2021). Further, humans have been observed to positively perceive emotional support provided by AI-enabled digital humans (Gelbrich et al., 2021). However, the perceived warmth of a digital human might be the reason for the satisfaction with emotional support provided by digital humans (Gelbrich et al., 2021).

While this may surprise, it implies that digital humans have progressed beyond providing information and direction to providing emotional support (Gelbrich et al., 2021). It is reasonable to argue that deploying digital humans in the workplace can provide employees with more warmth and consistent emotional support than humans. However, one implication is whether digital humans' provision of emotional support will change the human experience in the future. Further research is required to understand and apply human individuality in adapting to emotional support provided by digital humans (Fan et al., 2021; Matthews et al., 2021; Nawahdah & Inoue, 2013; Tan et al., 2021).

Moreover, one ethical concern is that modelling digital humans in terms of face and other appearance characteristics may give preferences to certain human characteristics. This could cause tension between natural and digital humans. Besides, while developers can design digital human models that resemble various groups of people, humans may perceive such attempts at diversity in digital human models as superficial.

**To provide information and guidance.** Businesses can use digital humans to deliver customer support and coaching while maintaining a human-to-human emotional connection (Brill et al., 2019; Tan et al., 2021). On the other hand, their use is not only to address labour/skill shortages but also to meet the growing demand of customers (Gelbrich et al., 2021; Tan et al., 2021). However, discussions about employing digital humans to provide information and guidance have generally assumed that humans will interact with digital humans similarly (Large et al., 2017). This is not the case since research is needed to assess, understand, and apply individual differences to digital/natural human interactions (Matthews et al., 2021).

Digital humans are expected to contextualise complex requests filled with cautionary words and expressions and vague language and utterances that involve shared contextual information and understanding (Fan et al., 2021; Large et al., 2017). Digital humans providing information and guidance should be in the 'right' virtual place to effectively guide humans (Nawahdah & Inoue, 2013). Therefore, a smile and a face are still insufficient for digital humans when providing information and guidance to natural humans.

**To enhance digital/natural human collaboration.** While the (natural) human brain can comprehend abstract concepts, digital humans' precision and speed significantly outpace human capabilities (Ghosh, 2002; Large et al., 2017). Such insight suggests that digital and natural humans can augment each other's strengths (Ghosh, 2002).

Using digital humans to research human factor issues (Dunston & Wang, 2005; Kuo & Chu, 2005; Maurice et al., 2017) in design and tasks, especially when certain tasks are risky for humans, may enhance digital/natural human collaboration (Fan et al., 2021; Kuo & Wang, 2009). Therefore, it is projected that digital humans would have implications for human existence (Guadagno et al., 2011; Matthews et al., 2021; Nowlan & Blake, 2007). On the other hand, such implications are geared toward assessing and understanding individual real-life human differences while interacting with digital humans (De Magistris et al., 2013; Matthews et al., 2021). While the literature can provide such insights, we have yet to realise the capabilities of digital humans fully (Khalid et al., 2018; Munoz et al., 2016), and so digital/natural human collaboration remains a futuristic topic (Jagdev et al., 2008).

**To protect the well-being of humans.** Because of the COVID-19 outbreak, debates about the deployment of digital humans to protect the well-being of humans appear to have gained traction (Díaz et al., 2021; Leng et al., 2021). The debates (Hu et al., 2011; Leng et al., 2021; Maurice et al., 2017) suggest that face-to-face interactions are limited during a pandemic like COVID-19. Some industries, such as telecommunication, need the human element as part of their operations and customer experience. Digital humans are suggested to provide the human element (Leng et al., 2021).

Previously, using digital humans to safeguard the health and well-being of humans was centred on lowering musculoskeletal disorders (MSDs) in the workplace (Hu et al., 2011; Maurice et al., 2017). While such uses of digital humans will continue to attract research and discussion, research into the use of digital humans to protect the well-being of humans will evoke further interest, whether these discussions are about protecting humans' well-being or merely attempting to reduce the human element in operations.

*4.2 Limitations of digital humans*

**Not correctly reproducing humans.** Digital humans may only capture the human-related features that are captured by motion and performance detection technologies (Guadagno et al., 2011; Kuo & Wang, 2009). Developing digital humans that do not accurately reproduce humans may not be an issue in some sectors, such as gaming and movie production; however, it is an issue when digital humans are used for evaluating products or workplaces to eliminate stressful work postures for humans (Harari et al., 2021; L. Ma et al., 2011; Maurice et al., 2017; Sanjog et al., 2019; Xiang et al., 2018).

Attempts to solve this issue include the use of medical photographs of human body parts (Harih et al., 2021; Lim et al., 2004; Zhang et al., 2010), the collection of demographic features of workers (Arunachalam et al., 2021; Y. Ma et al., 2011; Vyavahare & Kallurkar, 2016; Zhou et al., 2016), and the use of humans to conduct in-situ simulations (Harari et al., 2020; Harih & Dolšak, 2013; Summerskill et al., 2016). Despite these efforts, the digital reproduction of humans may overlook the symbolic abstract thinking, unpredictable nature and self-reflective consciousness of humans (Quarles et al., 2010).

While measuring human characteristics is utilised to build digital humans, the human reception of digital humans may also refine those measurements. As a result, digital humans may even be pushed beyond accurately reproducing humans (Choi et al., 2009; Chu et al., 2006; Harih et al., 2021).

**Digital human design decision issues.** Design choices are difficult for digital humans (Kim & Kim, 2020; Ye et al., 2018). One difficulty in enhancing ergonomics with digital humans, for example, is determining how to create them for the target demography (Kim & Kim, 2020; Ye et al., 2018). While this issue mandates demographically relevant data, it remains unclear what data should be. Using historical anthropometric data is one example of dealing with this issue (Choi et al., 2009; Harari et al., 2020; Y. Ma et al., 2011; Vyavahare & Kallurkar, 2016). In studies that used this approach, the target demographic's height and weight were generally used (Jung et al., 2009).

However, this approach has drawbacks to creating digital humans that represent a variety of demographics and generating digital humans for different target markets. One may also argue that the 'weight' of the intended target group can shift over time. While this attempt to use the target population's normal distribution of stature and weight will allow the creation of digital humans (Chander & Cavatorta, 2018; Jung et al., 2009), if these digital humans are used to model a tool, they will be limited in their ability to predict the tool user's force exertion, for example (Chander & Cavatorta, 2018; Harih et al., 2021; L. Ma et al., 2011). If historical data is used, the digital humans might have been created using out-of-date data, meaning they reproduce past target demographics.

Other attempts to address this issue include employing artificial neural networks (Zhang et al., 2010) to record human body postures. These efforts, however, rely on artificial intelligence, which requires further development (Gragg et al., 2013; Lämkull et al., 2009; Nawahdah & Inoue, 2013). Such advancements are required for artificial intelligence to accurately predict the outcome of a task, such as hand access, push and pull forces, leaning and balance behaviour, and field of vision (Harih & Dolšak, 2013; Lämkull et al., 2009; Nawahdah & Inoue, 2013). Therefore, design decisions may limit the performance of digital humans (Chu et al., 2006; Kuo & Wang, 2009; Sanjog et al., 2019).

**Not representing the target human population.** One of the limitations of digital humans is where the data about humans comes from. Existing approaches to developing digital humans either pool data from multiple populations (Harih et al.,

2021) or develop a group of digital humans using data from a single target population (Jung et al., 2009). The developed digital humans, in both instances, are limited in their ability to reflect the target human population.

Initiatives to deal with this issue include gathering data from volunteers (Arunachalam et al., 2021; Emmatty et al., 2021), female and male patients (Potvin & Potvin, 2019), and young and old drivers (Chateauroux & Wang, 2010; Large et al., 2017; Peng et al., 2017); nonetheless, effectively developing digital humans to represent various demographics remains a challenge (Maurice et al., 2017).

**Digital/natural human interaction issues.** When gesture, text and voice interactions are compared, natural humans least preferred the gesture interaction with digital humans (Silva & Bonetti, 2021). However, this study was conducted during the Covid-19 pandemic. Natural humans have had limited interactions among themselves.

While digital humans have the potential to reduce the 'natural' human element in industries such as fashion (Silva & Bonetti, 2021), if natural humans as customers of a brand, for example, do not approve of the form of interaction with digital humans, a business may be better off by keeping its natural human workforce to keep its customers.

Further, natural humans receive digitally rendered smiles positively (Guadagno et al., 2011). Therefore, digital humans' text and speech interactions are insufficient for natural humans (Chu et al., 2006; Guadagno et al., 2011). The interaction between digital and natural humans must thus be as life-like as technology permits (Large et al., 2017; Silva & Bonetti, 2021).

However, what a natural human's preferences will be when it comes to digital humans is still evolving (Large et al., 2017; Silva & Bonetti, 2021). While the focus of the discussion on digital/natural human interactions so far has been on digital human abilities, it is also critical to consider cultural characteristics of how a digital human should look or act (Guadagno et al., 2011). Further research is required on how digital humans are perceived in human culture. Cultural differences, such as how conservative or liberal society is, may influence how digital humans are received.

**Lack of universal documentation.** There is still a lack of universal documentation and no regulatory body to validate digital human documentation (Hanson et al., 2006; Harih et al., 2021). In this context, the term "universal" refers to a documentation system that is accessible to the public, whether they are businesses or individuals. There have been attempts toward such documentation systems. These efforts, however, are directed by organisations that work on digital humans (Hanson et al., 2006).

There have also been attempts to model digital humans (Arunachalam et al., 2021; Harih et al., 2021; Vyavahare & Kallurkar, 2016). These efforts include considering different target populations and age groups (Harih et al., 2021; Peng et al., 2017; Vyavahare & Kallurkar, 2016). These efforts, however, are founded on mathematics and philosophy (Wong et al., 2012).

**Risks and ethics of employing digital humans.** In evaluating the ergonomics of the workplace, digital humans have been used (Dunston & Wang, 2005; Sanjog et al., 2019). We may nevertheless place significant faith in the technology's performance in developing the initial prototypes (Kim & Kim, 2020; Tan et al., 2021).

Because of the technology's convenience, there may be assumptions about ergonomics for humans outside of the trials (Demirel et al., 2021; Kim & Kim, 2020; Tan et al., 2021; Zhou et al., 2016).

Moreover, their deployment may also have negative employment consequences and risks to safety, privacy, and liability (Burleigh et al., 2013; Kim & Kim, 2020). Humans are still unwilling to delegate to digital humans, and delegated tasks to digital humans are generally viewed with scepticism (Gogoll & Uhl, 2018). The perceived usefulness of digital humans, such as in care homes, is mitigated by the 'trust' that humans have in this technology (Gogoll & Uhl, 2018; Tan et al., 2021).

Therefore, it is still uncertain whether humans can trust digital humans with people (Tan et al., 2021). While this may be due to humans' 'trust' in the level of the technology (Gogoll & Uhl, 2018), there are also ethical concerns such as loss of freedom, loss of human interaction and social connection, dehumanisation and fraud (Tan et al., 2021).

**Human perception of digital humans.** The social evaluation of digital humans limits their development and deployment in organisations (Guadagno et al., 2011). In a Western-like workplace, a female digital human without a head covering can be accepted; however, in a conservative society, the same digital human may be socially rejected (Ranger et al., 2019). While there are concerns about digital humans' societal acceptance, we have not grasped their social significance.

A digitally rendered smile or encouraging words from digital humans may be comforting in a care home, but it is unclear whether these elements are socially valuable and can be emotional bonds (Burleigh et al., 2013; Guadagno et al., 2011; Large et al., 2017). In this instance, perceiving a digitally rendered smile as positive by natural humans may be because humans are still tolerant of digital humans' limitations (Yam et al., 2020).

**Costly and time-consuming development.** Developing digital humans is resource-intensive (Lee et al., 2011; Ranger et al., 2019) and takes time and financial resources (Lee et al., 2011). Even if an organisation devotes time and financial resources to developing digital humans, the issue of whether the organisation can find skilled human resources to develop digital humans remains (Lee et al., 2011). Furthermore, choosing the "right" digital human model will further deplete resources (Ranger et al., 2019). Therefore, developing digital humans is costly and complex (Ranger et al., 2019). Given the uncertainty around the return on investment, organisations may perceive resources used for that purpose to be wasted (Lee et al., 2011; Ranger et al., 2019).

**Being prone to cyber-attacks.** Digital humans can be controlled and are vulnerable to cyber-attacks because of their digital forms (Khalid et al., 2018). Due to their digital forms, they can also be reproduced into as many copies as desired in cyberspace (dos Santos et al., 2021).

     Cyber-attacks on digital humans, particularly when humans are entrusted in their care, have serious ethical and legal ramifications beyond trust in digital/natural human collaboration (Khalid et al., 2018). This includes the use of digital humans for financial scamming. A digital replica of a known person might be used to persuade people to send money to an address in exchange for a doubled amount of their money. Cyberattacks of this nature are becoming increasingly sophisticated (Wong et al., 2012).

**5. Concluding remarks**

'Digital humans' are still a future technological breakthrough. While advances in constructing human-like digital beings have been made, such advancements are still in their formative stages. Existing digital human prototypes have been trained to reproduce a variety of human tasks, such as assisting, providing support, and so on.

The literature around digital humans primarily discusses developing digital human models and employing digital humans to enhance ergonomics. However, additional value propositions can be observed in the existing literature. These include alleviating labour/skill shortages, decreasing the human element in operations, enhancing natural/digital human collaboration, giving information and direction, providing emotional support, evaluating products, lowering costs, and saving time in product development.

However, given the limits of digital humans and factors such as developing resources and ethical and legal concerns, these value propositions have yet to be fully realised.

**5.1 Future directions**

One central line of research on digital humans is developing and using digital human models to enhance design ergonomics (Chaffin, 2005, 2009). Other streams of literature, on the other hand, have begun but are still underexplored, such as the use of digital humans to alleviate labour/skill shortages (Tan et al., 2021; Xiang et al., 2018) and the consequences of deploying digital humans on human employment prospects (Dekker et al., 2017).

Existing research (Kim & Kim, 2020; Tan et al., 2021; Xiang et al., 2018) suggests that digital humans can alleviate labour/skill shortages. Future research can contribute to the practicality, ethicality, and legality aspects of deploying digital humans to alleviate labour/skill shortages in the workplace (Burleigh et al., 2013; de Kerckhove, 2021; Gogoll & Uhl, 2018; Tan et al., 2021).

Also, a paradox exists about developing and deploying digital humans in the literature. On the one hand, creating digital humans is resource-intensive (Y. Ma et al., 2011; Ranger et al., 2019). On the other hand, using digital humans is pitched to lower labour costs (Caputo et al., 2019; Jagdev et al., 2008; Xiang et al., 2018). Future research can investigate this paradox to determine if it is a 'both/and' choice (Smith & Lewis, 2011) for organisations to develop and deploy digital humans.

Future projections suggest that digital humans may outperform humans in human-only skills (Holford, 2019; Sousa & Wilks, 2018). Further research can investigate the implications of deploying digital humans and how they may affect human job prospects (Giudice et al., 2021; Yam et al., 2020). It is also compelling to continually consider how digital/natural humans can coexist in the workplace in the future (Saracco et al., 2021; Wilson & Daugherty, 2019).

While the literature (Gelbrich et al., 2021; Guadagno et al., 2011; Yam et al., 2020) suggests that humans consider digital smiles favourable, the existing deployment of digital humans undermines this (Dekker et al., 2017; Tan et al., 2021). This positive perception might be because humans are tolerant of digital humans' limitations (Yam et al., 2020). It is unclear whether this favourable view will persist when, for example, cultural factors are considered. Further research can explore the social significance and acceptance of digital humans in the workplace across cultures. This line of research can extend beyond natural humans' perceptions of digital humans (Guadagno et al., 2011) to include human model design decision-making issues (Jung et al., 2009; Y. Ma et al., 2011), universal documentation (Ye et al., 2018), and digital representations of various groups of people (Arunachalam et al., 2021; Potvin & Potvin, 2019).

The diversity of digital humans will be an interesting future area of research (Jung et al., 2009; Vyavahare & Kallurkar, 2016). A selective approach has been adopted to develop existing digital human models representing humans (Harari et al., 2020; Y. Ma et al., 2011). Being selective in developing digital humans tends to disregard the diverse physical characteristics we see in real people. Existing models reproduce a human model's physical features (Y. Ma et al., 2011; Vyavahare & Kallurkar, 2016). However, this may have social and ethical implications, such as assigning preference to certain human physical characteristics, whether intentionally or unintentionally. Humans with certain physical characteristics may feel superior or inferior. Future research might investigate how to improve the diversity of digital humans.

## 5.2 Contribution

This review suggests that the perceived usefulness and ease of development of digital humans determine the willingness of organisations to utilise this technology (Davis, 1989; Davis et al., 1989). The current digital human offerings are promising but limited in usefulness and ease of development. In practice, workers and their organisations can benefit from digital humans. However, before making any development and deployment decisions, organisations must assess the technology's usefulness when considering its implications and resource-intensive nature.

## 5.3 Limitations

Some limitations will be indicated. The review is limited to CABS-ranked journal articles, with non-CABS, non-peer-reviewed papers, books and book chapters, and practitioner research being excluded. Second, the search string was created to retrieve relevant articles from the Web of Science (WoS). This search string may not have returned other relevant journal articles outside the Web of Science database. Third, while the keywords were obtained from published articles, they might not have returned all relevant articles. While acknowledging these limitations, this paper aims to encourage future research into the value and limitations of digital humans in organisations.